\documentclass[epj]{svjour}
\usepackage{graphics}
\usepackage[figuresright]{rotating}
\begin{document}

\title{
%\vskip 1.2cm}
Quark Number Susceptibilities \& The Wr\'oblewski Parameter }
\author{Rajiv V. Gavai\inst{1,2}
\thanks{\emph{Alexander von Humboldt Fellow on leave from TIFR, Mumbai 
and Speaker at the Conference.}}\and 
Sourendu Gupta\inst{1} 
}                     % Do not remove
\institute{Department of Theoretical Physics, Tata Institute of Fundamental
         Research,\\ Homi Bhabha Road, Mumbai 400005, India.
\and Fakult\"at f\"ur Physik, Universit\"at Bielefeld, 
         D-33615, Germany.}
\date{Report No. : TIFR/TH/05-08, hep-ph/0502198} 
\abstract{
The Wr\'oblewski parameter is a convenient indicator of strangeness production
and can be employed to monitor a signal of quark-gluon plasma production :
enhancement of strangeness production.  It has been shown to be about a factor
two higher in heavy ion collisions than in hadronic collisions.   Using a
method proposed by us earlier, we obtained lattice QCD results for the
Wr\'oblewski parameter from our simulations of QCD with two light quarks both
below and above the chiral transition.  Our first principles based and
parameter free result compare well with the A-A data from SPS and RHIC.
 % end of PACS codes
} %end of abstract
\maketitle
\section{Introduction}
\label{intro}

As with many signals of quark-gluon plasma (QGP) production in relativistic
heavy ion collisions, the basic idea behind enhancement of strangeness
production \cite{RM} as a QGP signal is very simple.  Recognising the fact that
the strange quark mass  is smaller than the expected transition temperature
whereas the mass of the lowest strange hadron is significantly larger, it was
argued that the production rate for strangeness in the QGP phase, $\sigma_{QGP}
(s \bar s)$ is greater than that in the hadron gas phase, $ \sigma_{HG}(s \bar
s)$.  While this energy threshold argument for strangeness production in the
two phases is qualitatively appealing, one has to face quantitative questions
of details for any meaningful comparison with the data.  Applications of
perturbative QCD needs a large scale which could be either the temperature of
QGP or the mass of the produced strange quark-antiquark pair.  Since the
temperature of the plasma produced in RHIC, or even LHC, may not be
sufficiently high for perturbative QCD to be applicable and since the strange
quark mass is also rather low , estimation of strangeness production by lowest
order processes like $gg \to s \bar s$ could be misleading.  Indeed, it is now
well-known that even for the charm production, the next order correction to $gg
\to c \bar c$ is as large as the leading order; such an order by order
computational approach may be hopelessly futile for the lot lighter strange
quark.

A variety of aspects of the strangeness enhancement have been studied and 
many different variations have been proposed.   One very useful way of
looking for strangeness enhancement is the Wr\'oblewski parameter \cite{Wr}.
Defined as the ratio of newly created strange quarks to light quarks, 
\begin{equation}
\lambda_s = \frac{2\langle s\bar s\rangle}{\langle u\bar u+d\bar d\rangle}
\end{equation}
the Wr\'oblewski parameter has been estimated for many processes using a hadron
gas fireball model \cite{BH}.  An interesting finding from these analyses is
that $\lambda_s$ is around 0.2 in most processes, including proton-proton
scattering, but is about a factor of two higher in heavy ion collisions. An
obvious question one can ask is whether this rise by a factor of two can be
attributed to the strangeness enhancement due to quark gluon plasma and if yes,
whether this can be quantitatively demonstrated by explicitly evaluating the
Wr\'oblewski parameter in both phases.  Alternatively, one could just study how
different the prediction actually is and learn about other physics effects from
its comparison with data.  We show below how quark number susceptibilities,
obtained from simulations of lattice QCD, may be useful in answering such 
questions.  Since these simulations correspond to equilibrium situations, one
needs certain extra assumptions which we also discuss briefly.

%%%%%%%%%%%%%%%%%%%%%%%%%%%%%%%%%%%%%%%%%%%%%%%%%%%%%%%%%%%%%%%%%%%%%%%%

\section{$\lambda_s$ from Quark Number Susceptibilities}

%%%%%%%%%%%%%%%%%%%%%%%%%%%%%%%%%%%%%%%%%%%%%%%%%%%%%%%%%%%%%%%%%%%%

%%%%%%%%%%%%%%%%%%%%%%%%%%%%%%%%%%%%%%%%%%%%%%%%%%%%%%%%%%%%%%%%%%%%
\begin{figure*}
\begin{center}
\includegraphics[scale=0.8]{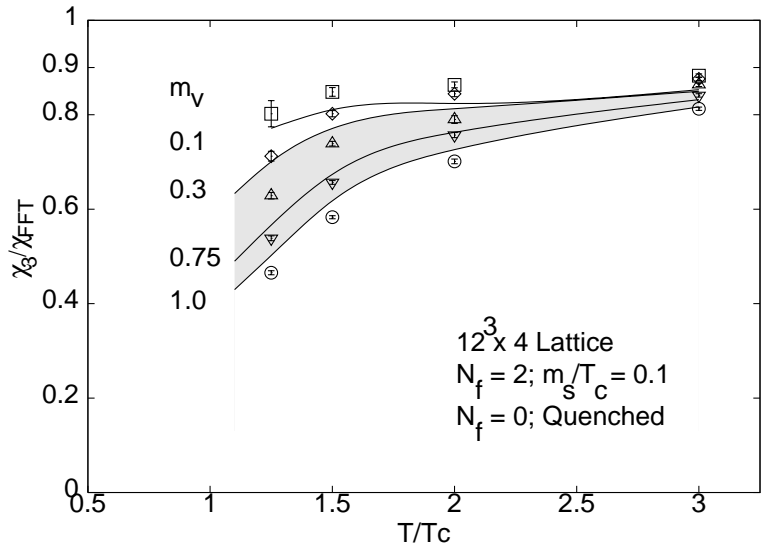}
\end{center}
\caption{Comparison of quenched and full QCD results.}
\label{fig1a}
\end{figure*}
\begin{figure*}
\begin{center}
\includegraphics[scale=0.8]{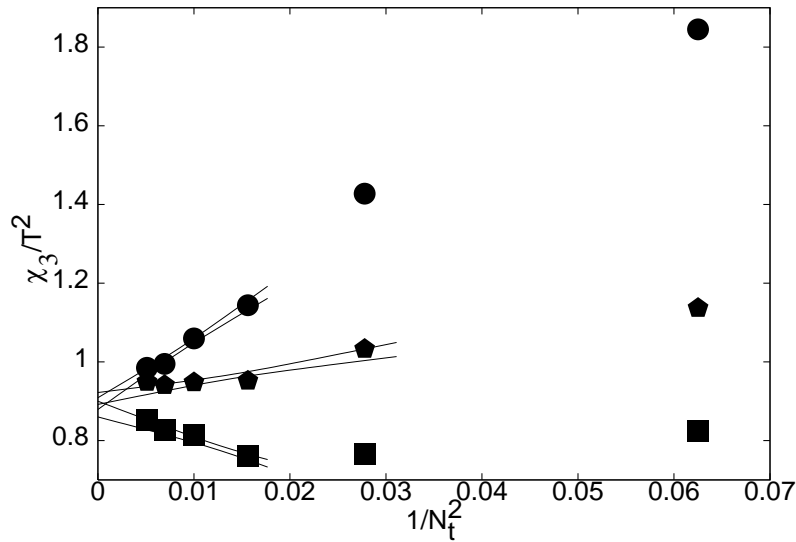}
\end{center}
\label{fig1b}
\caption{Typical continuum extrapolation results.}
\end{figure*}

%%%%%%%%%%%%%%%%%%%%%%%%%%%%%%%%%%%%%%%%%%%%%%%%%%%%%%%%%%%%%%%%%%%

%%%%%%%%%%%%%%%%%%%%%%%%%%%%%%%%%%%%%%%%%%%%%%%%%%%%%%%%%%%%%%%%%%%%
\begin{figure*}
\begin{center}
\includegraphics{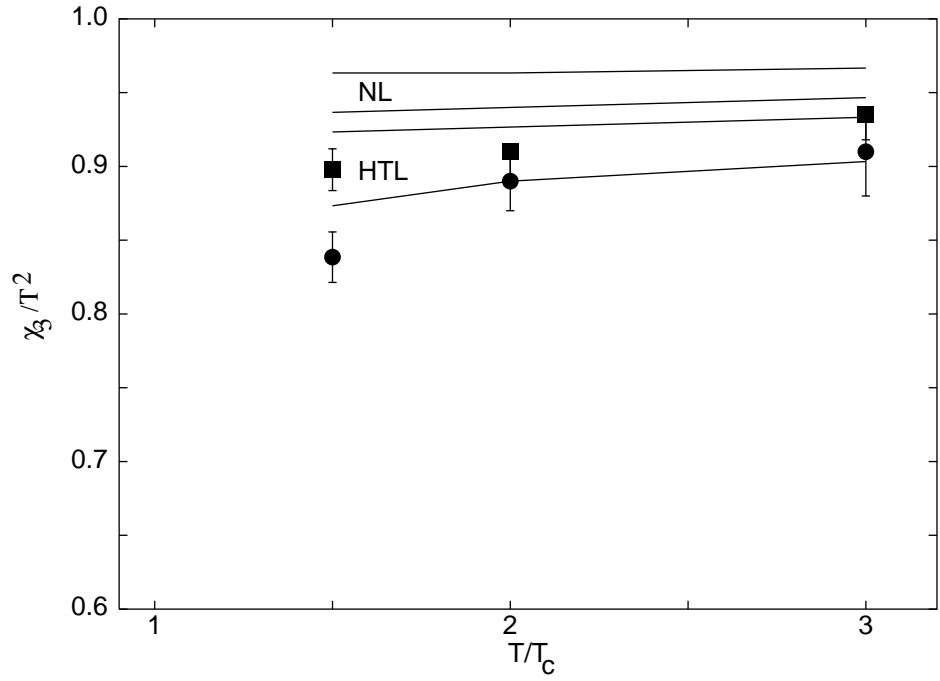}
\end{center}
\caption{Quenched QCD results in continuum limit.} 
\label{fig2a}
\end{figure*}
\begin{figure*}
\begin{center}
\includegraphics[width=12cm]{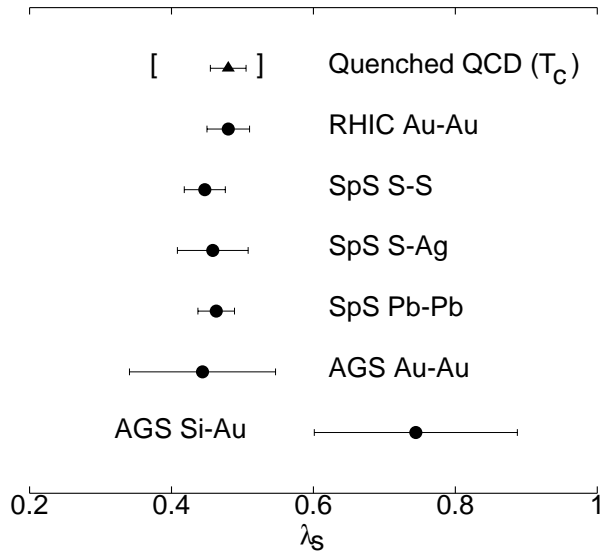}
\end{center}
\caption{Comparison of the corresponding $\lambda_s$ with RHIC and 
SPS experiments.}
\label{fig2b}
\end{figure*}

%%%%%%%%%%%%%%%%%%%%%%%%%%%%%%%%%%%%%%%%%%%%%%%%%%%%%%%%%%%%%%%%%%%%%%%%

%%%%%%%%%%%%%%%%%%%%%%%%%%%%%%%%%%%%%%%%%%%%%%%%%%%%%%%%%%%%%%%%%%%%
\begin{figure*}[htbp]\begin{center}
\includegraphics[angle=270,width=12cm] {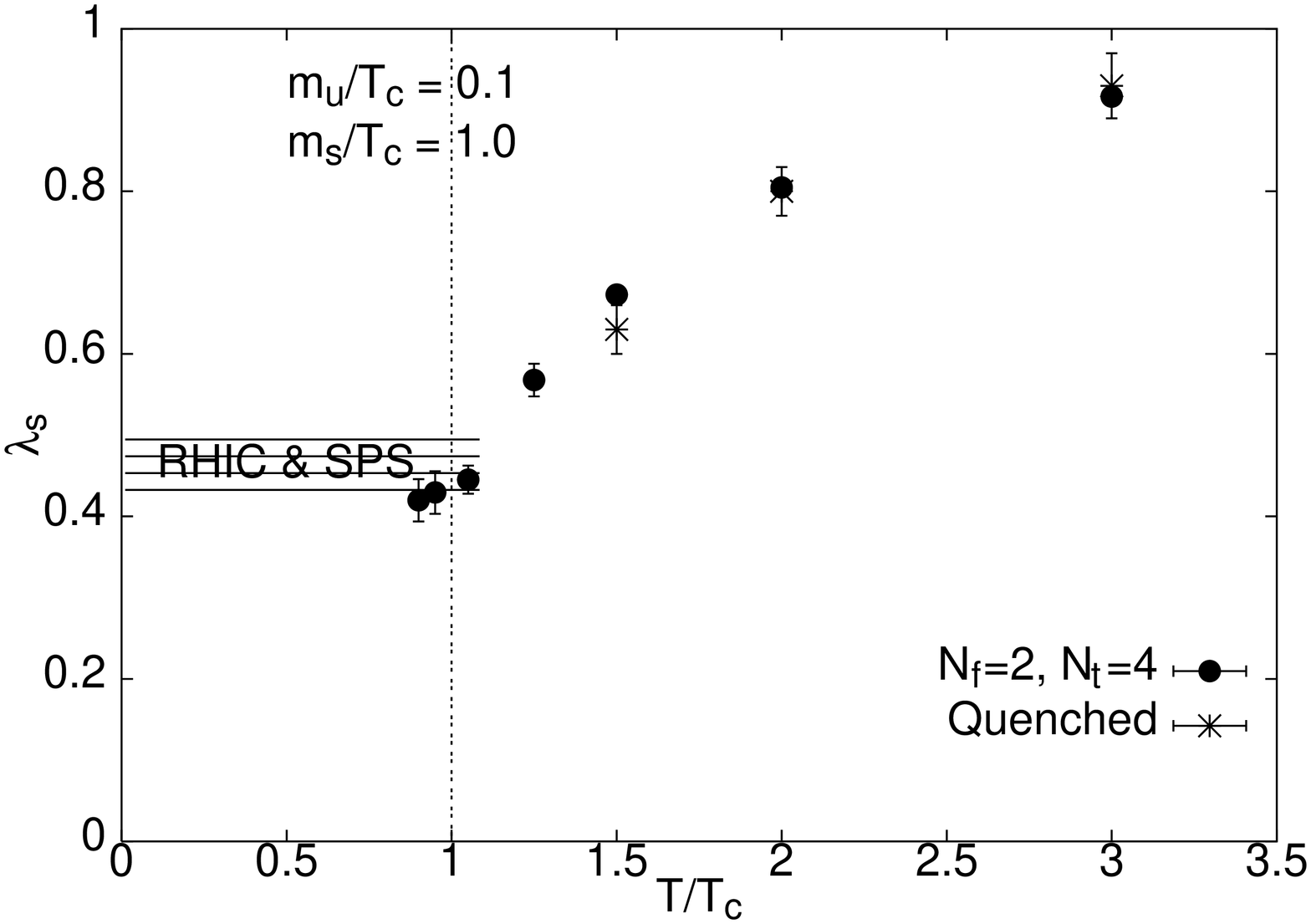}
\caption{$\lambda_s$ as a function of temperature for full and quenched QCD.}
\label{fig3}\end{center}
\end{figure*}
%%%%%%%%%%%%%%%%%%%%%%%%%%%%%%%%%%%%%%%%%%%%%%%%%%%%%%%%%%%%%%%%%%%

Quark number susceptibilities (QNS) can be calculated from first principles
using the lattice formulation. Assuming three flavours, $u$, $d$, and $s$
quarks, and denoting by $\mu_f$ the corresponding chemical potentials, the QCD
partition function is
\begin{equation}
{\cal Z} =  \int D U~~ \exp(-S_G)  \prod_{f=u,d,s} 
{\rm Det}~M(m_f, \mu_f)~~.
\label{zqcd}
\end{equation}
Note that the quark mass and the corresponding chemical potential enter only
through the Dirac matrix $M$ for each flavour.  We use staggered fermions
and the usual fourth root trick \cite{review} to define $M$ for each flavour.
Defining $\mu_0 = \mu_u + \mu_d + \mu_s$ and $\mu_3 = \mu_u - \mu_d$, the
baryon and isospin densities and the corresponding susceptibilities can be
obtained as:  

\begin{equation}
\qquad \qquad n_i =  \frac{T}{V} {{\partial \ln {\cal Z}}\over{\partial \mu_i}}, \qquad
\chi_{ij} =  \frac{T}{V} {{\partial^2 \ln {\cal Z}}\over{\partial \mu_i 
\partial \mu_j} }  ~~.
\label{nchi}
\end{equation}

QNS in eq. (\ref{nchi}) are crucial for many quark-gluon plasma signatures which
are based on fluctuations in globally conserved quantities such as baryon
number or electric charge.  Theoretically, they serve as an important
independent check on the methods and/or models which aim to explain the large
deviations of the lattice results for pressure $P$($\mu$=0) from the
corresponding perturbative expansion.  Here we will be concerned with the
Wr\'oblewski parameter which we \cite{us} have argued can be estimated from the
quark number susceptibilities:
\begin{equation}
\lambda_s = {2 \chi_s \over { \chi_u +\chi_d}}~. 
\label{wrob}
\end{equation}

Note that the lattice simulations yield real quark number susceptibility
whereas for particle production its imaginary counterpart is needed.  Indeed,
$\lambda_s$ above too needs the latter. However, one can relate the two and thus
justify the use of lattice results in obtaining $\lambda_s$.  Briefly, the
argument \cite{book} is as follows. Fluctuations in physical quantities,
described by a perturbation in time, can be related to a generalized
susceptibility for the corresponding operator for it.  This is complex in
general.  Its imaginary part can be shown to determine the dissipation,  
i.e., the production of strange quark-antiquark pair in our case.  From the
general properties of these susceptibilities, a Kramers-Kronig type relation 
between their real and imaginary parts can be obtained.  Finally, making a relaxation time approximation ($\omega \tau \gg 1$), one finds that the
ratio of the imaginary parts is the same as that of the real parts.

In order to use eq. (\ref{wrob}) to obtain an estimate for comparison with
experiments, one needs to compute the corresponding quark number
susceptibilities on the lattice first and then take the continuum limit.  All
susceptibilities can be written as traces of products of the quark propagator,
$M^{-1}(m_q)$, and various derivatives of $M$ with respect to $\mu$.  With $m_u
= m_d$, diagonal $\chi_{ii}$'s can be written as

\begin{eqnarray}
\label{chiexp1}
\chi_0 &=& \frac{T}{2V} [ \langle {\cal O}_2(m_u) + \frac{1}{2} {\cal O}_{11}(m_u) \rangle ] \\ 
\label{chiexp2}
\chi_3 &=& \frac{T}{2V} ~~ \langle {\cal O}_2(m_u) \rangle  \\ 
\label{chiexp3}
\chi_s &=& \frac{T}{4V} [ \langle {\cal O}_2(m_s) + \frac{1}{4} {\cal O}_{11}(m_s) \rangle ] ~~.
\end{eqnarray} 

\noindent
Here ${\cal O}_2 = {\rm Tr}~M^{-1}_u M_u'' - {\rm Tr} ~M^{-1}_u M_u'M^{-1}_u
M_u'$, and $ {\cal O}_{11}(m_u) = ({\rm Tr}~M^{-1}_u M_u' )^2$.  The traces are
estimated by a stochastic method: $ {\rm Tr}~A = \sum^{N_v}_{i=1} R_i^\dag A
R_i / 2N_v$, and $ ({\rm Tr}~A)^2 = 2 \sum^{L}_{i>j=1} ({\rm Tr}~A)_i ({\rm
Tr}~A)_j/ L(L-1)$, where $R_i$ is a complex vector from a set of $N_v$,
subdivided further in L independent sets. We use typically $N_v =$ 50-100.

Figure 1 displays results \cite{us1} for the susceptibilities as a function of
temperature in units of $T_c$, where $T_c$ is the transition temperature.
Normalized to the corresponding ideal gas results on the same lattice, i.e, the
infinite temperature limit, results for QCD with two light dynamical quarks of
mass $0.1~T_c$ are shown as points whereas the continuous curves correspond to
the results in the quenched approximation.  Note that the latter amounts to 
dropping the fermion determinant term in simulations which become orders
of magnitude faster, and hence more precise, than the full QCD simulations.
The valence quark mass $m_{\rm v}$, appearing in eqs.
(\ref{chiexp1})-(\ref{chiexp3}), is shown in the figure in units of $T_c$.
Note that $T_c$ in these two cases differ by a factor of 1.6 but the results
for the corresponding dimensionless susceptibilities as a function of the
dimensionless ratio $T/T_c$ differ by a few per cent only.  Such a mild
dependence on the number of dynamical flavours in the thermodynamic quantities
has been a known feature in the temperature region away from the transition.
Indeed, since the nature of the transition does depend strongly on the number
of dynamical flavours, one expects significant differences near $T_c$.
Encouraged by this behaviour, we investigated the continuum limit for the
quenched case by increasing the temporal lattice size from 4 to 14 in steps of
two and extrapolating to infinite temporal lattices.  The spatial lattices were
also increased to maintain the aspect ratio constant.   Figure 2 shows typical
results of such continuum extrapolation at $T = 2~T_c$.  The continuum results
for the light quark susceptibility thus obtained in the quenched approximation
are exhibited in Figure 3 for small $m_{\rm v}$. The bands marked by `HTL' and
`NL' show the analytic results of \cite {toni} obtained in successively 
better approximations respectively. 

The strange quark susceptibility in the continuum limit was obtained from the
same simulations by simply choosing $m_{\rm v}/T_c = 1$ (in both full and
quenched QCD in view of Figure 1. Using eq. (\ref{wrob}), $\lambda_s(T)$ can
then be easily obtained.  These were extrapolated to $T_c$ by employing simple
ans\"atze.  The resultant $\lambda_s(T_c)$ in quenched QCD is shown in Figure 4
along with the results obtained from the analysis of the RHIC and SPS data in
the fireball model\cite{BH}.  The systematic error coming from extrapolation is
shown by the brackets.  The agreement of the lattice results with those 
from RHIC and SPS is indeed very impressive.

The nice agreement needs to be treated cautiously, however, in view of the
various approximations made. Let us list them in order of severity.

\begin{itemize}

\item The result is based on quenched QCD simulations and extrapolation to
$T_c$.  As seen from Figure 1, the quark number susceptibilities, and hence
$\lambda_s(T)$, are expected to change by only a few per cent.  Since the
nature of the phase transition does depend strongly on the number of dynamical
quarks, a direct computation near $T_c$ for full QCD is desirable.  We  are
currently making such a computation and have some preliminary results for full
QCD with two light dynamical quarks for lattices with four sites in temporal
direction.  These are shown in Figure 5 along with the continuum quenched
results for $\lambda_s(T)$ and the band for experimental results.  While the
emerging trend is encouraging, further exploration with varying strange quark
mass, temporal lattice size (to obtain continuum results) and spatial volume is
still necessary. 

\item The experiments at RHIC and SPS have nonzero albeit small $\mu$ whereas
the above result used $\mu=0$. Based on both lattice QCD and fireball model
considerations, $\lambda_s$ is expected to change very slowly for 
small $\mu$.  This can, and should, be checked by direct simulations.

\item As argued above, particle production needs the imaginary counterpart of
what one obtains from simulations. The relation between the ratios of real and
imaginary parts was obtained under the assumption that the characteristic time
scale of quark-gluon plasma are far from the energy scales of strange or light
quark production. Observation of spikes in photon production may falsify this
assumption.

\end{itemize}
%%%%%%%%%%%%%%%%%%%%%%%%%%%%%%%%%%%%%%%%%%%%%%%%%%%%%%%%%%%%%%%%%%%%%%%%%
\section{Summary}

Quark number susceptibilities contribute in many different ways to the physics
of the signals of quark-gluon plasma in heavy ion collisions at SPS and RHIC.
They can be obtained from first principles using lattice QCD.  This offers a
quantitative control and check of these signals and thus QGP itself.  In
particular, the continuum limit of $\chi_u$ and $\chi_s$, which we obtained
in quenched QCD, leads to a temperature dependent  Wr\'oblewski parameter,
$\lambda_s(T)$.  Its extrapolation to $T_c$ appears
to be in good agreement with results from SPS and RHIC. First full QCD
results near $T_c$ confirm this as well, although many technical issues,
e.g, finite lattice cut-off or strange quark mass, need to be sorted out still.

%%%%%%%%%%%%%%%%%%%%%%%%%%%%%%%%%%%%%%%%%%%%%%%%%%%%%%%%%%%%%%%%%%%%%%%%%
\section{Acknowledgements}

One of us (RVG) thanks the Alexander von Humboldt Foundation, Germany for the
financial support without which participation in this excellent and exciting 
meeting would not have been feasible.  It is also a pleasure to acknowledge 
the support of the organizers, especially Profs. Carlos Lourenco and Helmut 
Satz. 

\vspace{0.5cm}

%%%%%%%%%%%%%%%%%%%%%%%%%%%%%%%%%%%%%%%%%%%%%%%%%%%%%%%%%%%%%%%%%%%%%%%

\end{document}